# Search for Single Top Production at the Tevatron

A. Gresele
*Department of Physics, via Sommarive 14,
38050 Povo (Trento), Italy*

We report on a search for Standard Model t-channel ans s-channel single top quark production in $p\bar{p}$ collisions at a center of mass energy of 1.96 TeV. We use a data sample corresponding to 0.7 $fb^{-1}$ recorded by the upgraded Collider Detector at Fermilab (CDFII) and a data sample corresponding to 370 $pb^{-1}$ recorded by DØ. Both CDF and DØ find no significant evidence for electroweak top quark production and set upper limits at the 95% confidence level on the production cross section.

## 1 Single Top at the Tevatron

At the Tevatron, the most important production mode for top quarks is the strong interaction, top quarks are produced in pairs. Top quarks can also be produced via electroweak interactions, they are then produced singly[1]. The two relevant processes are the *t*- and the *s*-channel exchange of a virtual W boson. The cross section predicted by the Standard Model are (1.98 ± 0.25) pb and (0.88 ± 0.11) pb at $\sqrt{s}$ = 1.96 TeV respectively[2]. A measurement of these cross sections is of interest for the determination of $|V_{tb}|^2$, which could give a hint on the assumed unitarity of the CKM matrix.

## 2 CDF Single Top analysis

For this analysis, we consider only decays of the W boson into an electron or muon and the respective neutrino. The experimental signature is therefore an isolated electron or muon with $E_T$ respectively $p_T$ larger than 20 GeV, missing transverse energy larger than 20 GeV and a jet tagged as a b-jet. In the s-channel, another *b* quark can be measured as b-tagged jet, coming from the decay of the virtual W boson. In the t-channel, the *b* quark coming from the splitting of the gluon is rather forward. However, the t-channel has as an additional experimental signature a light quark jet.The most important backgrounds are $t\bar{t}$ quark productions, dilepton events, W+jets production and QCD multijet production faking an electron or muon and missing transverse energy. Table 1 summarizes the event yield obtained with a data sample corresponding to an integrated luminosity of 0.7 $fb^{-1}$ (only events in the W+2-jets-bin are considered). For both s+t combined and s-/t- channel separate searches CDF uses the Neural Network tecnique. These neural networks are three layer perceptrons, implemented using the NeuroBayes package[3]. There are therefore three distinct neural networks used in this analysis, two for the individual s- and t- signal channels and one for the combined s+t channel. In the training of the s-channel network, the t-channel events are treated as background and vice versa.

Table 1: Predicted and observed events after all selection requirements have been imposed.

|  | W + 2 jets |
|---|---|
| Total Background | 645.9 ± 96.1 |
| Single Top | 28.2 ± 2.6 |
| Total Prediction | 674.1 ± 96.1 |
| Observation | 689 |

## 2.1  s + t Combined Search Results

Based on the NN output shape similarity of the different MC distributions, we group the different physiscs processes into four classes: single top, $t\bar{t}$, charm like and bottom-like. The statistical method is fully Bayesian and it is described in the 2005 single top PRD article[4]. The Bayesian likelihood function consists of Poisson terms for the individual bins of the fitted histrograms. Systematic uncertainties are included as factors modifying the expectation value of events in a certain bin. The shape uncertainties are estimated from the shifted NN output histograms corresponding to MC samples in which the particular sources for systematic uncertainties are varied. Pseudo-experiments from the SM background+SM signal samples are generated and the Bayesian procedure are used to determine the most probable value of the signal cross section along with the 95% C.L. limits. Our expected limit correspond to the median of the indiviadual 95% C.L. limits distribution, which is 5.7 pb (Fig. 1, left plot).

In case of the CDF data, the likelihood fit yields a best value for the s+t cross section of $0.8^{+1.3}_{-0.8}(\text{stat.})^{+0.2}_{-0.3}(\text{syst.})$ pb. The resulting upper limit on the s+t cross section is 3.4 pb at 95% C.L. The posterior probability density is shown in the right plot of Fig. 1. The fit result is illustrated in Fig. 2 (left hand-side). The data and the expectation in the signal region are shown on the right hand-side of Fig. 2.

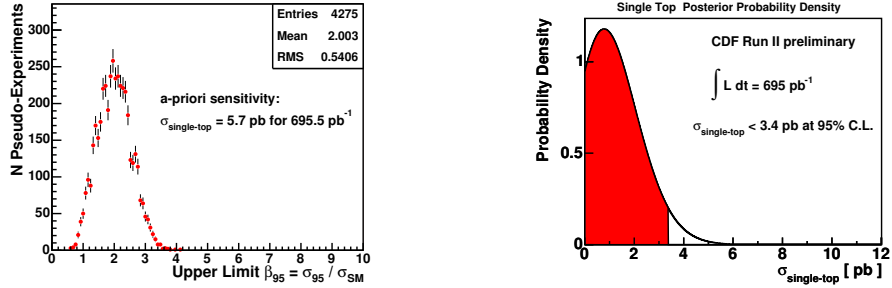

Figure 1: Expected distribution of 95% C.L. upper limits on $\sigma_{s+t}$ and observed posterior probability density.

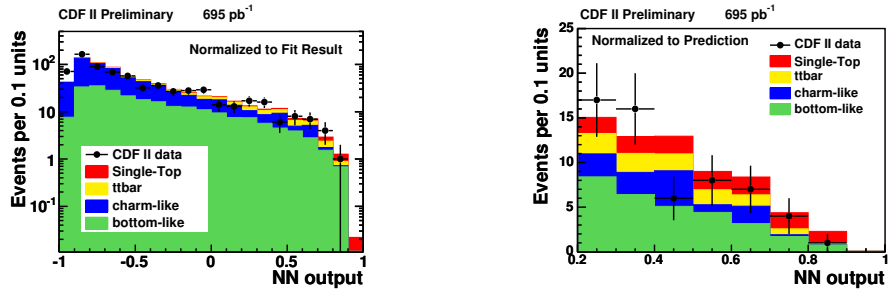

Figure 2: Data distribution of the s+t NN output in the entire output region and the signal region.

## 2.2 s- and t- Channel Separate Search

Two distinct neural netwrok are used, one trained for s-channel and the other one for the t-channel, which provide the opportunity to search for both channel individually and simultaneously. The training and input variables of the networks are similar to the one described in the previous section except that now each net is optimized for one (s- or t-) channel alone. Repeating the same procedure, pseudoexperiments using the SM signal contribtutions are permormed and fitted to the 2D templates. The expected 95% C.L. limits on the t- and s- cross section are 4.2 pb and 3.7 pb, respectively.

Using the CDF data, the likelihood fitting procedure returns $\sigma_{t-ch} = 0.6^{+1.9}_{-0.6}(\text{stat.})^{+0.1}_{-0.1}(\text{syst.})$ pb and $\sigma_{s-ch} = 0.3^{+2.2}_{-0.3}(\text{stat.})^{+0.5}_{-0.3}(\text{syst.})$ pb. This is graphically shown in Fig. 3. At the 95% C.L. the resulting upper limits on the t- and s-channel cross section are 3.1 pb and 3.2 pb, respectively.

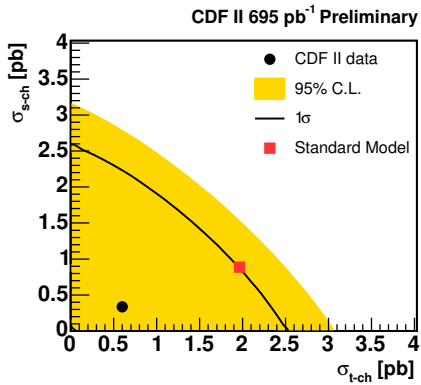

Figure 3: The separate search results, showing the most probable value (circle) the SM prediction (square) and the region excluded at 95% C.L. by the data (yellow).

## 3 DØ Single Top analysis

The analysis partitioned by decay mode, in electron and muon channels, is based on Run II data corresponding to an integrated luminosity of $366 \pm 24\,pb^{-1}$ for the electron channel and $363 \pm 24\,pb^{-1}$ for the muon channel. The events selected must contain only one isolated lepton with $p_T \geq 15$ GeV. In order to account for the presence of a neutrino, we require missing transverse energy larger than 15 GeV. The search is performed with one b-tagged jet and with at least 2 b-tagged jets. Table 2 summarizes the predicted and observed events after all selection requirements have been imposed.

Table 2: Predicted and observed events after all selection requirements have been imposed.

|                  | s-channel search | t-channel search |
|------------------|------------------|------------------|
| Total Background | 287.4 ± 31.4     | 275.8 ± 31.5     |
| Observed events  | 283              | 271              |

### 3.1 Likelihood Discriminant Method

After the event selection, a final discriminating likelihood variable is constructed in order to efficiently characterize the signal type events and reject the background type ones.We therefore set upper limits at the 95% C.L., using a Bayesian approach. As the main background ($t\bar{t}$ and W+jets) have different topologies, we choose to build two likelihood discriminants for each channel: a signal/$t\bar{t}$ filter and a signal/W+jets filter (see Fig. 4). Electron and muon channels in each tagging scheme are treated as indipendent channels, and are combined in the limit calculation, which leads to a total of 16 likelihood discriminant variables. We assume a Poisson distribution for the observed counts, and a flat prior probability for the signal cross section. The priors for the signal acceptance and the backgrounds are multivariate Gaussians centered on their estimates and described by a covariance error matrix taking into account correlations across the different sources and bins. We combined the four orthogonal analysis channels (electron and muon, single and double tags) to enhance the sensitivity of the analysis. Plots for the Bayesian posterior density are provided in Fig. 5. The observed (expected) 95% C.L. limits are 5.0 pb (3.3 pb) for the s-channel and 4.4 pb (4.3 pb) for the t-channel.

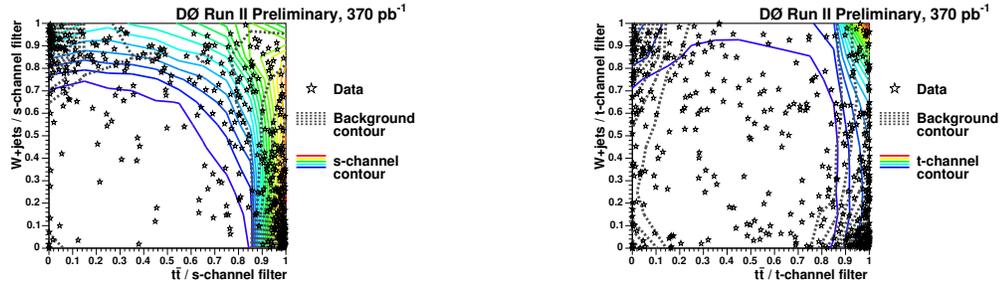

Figure 4: Expected distribution of 95% C.L. upper limits on $\sigma_{s+t}$ and observed posterior probability density (right) for the combined search using the s+ t NN output.

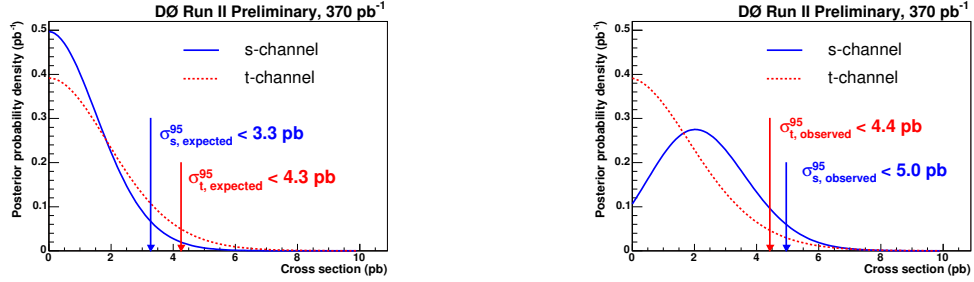

Figure 5: Data distribution of the s+t NN output in the entire output region (left hand-side) and the signal region (right hand-side).